# An Experimental Study of Load Balancing of OpenNebula Open-Source Cloud Computing Platform

A B M Moniruzzaman, *StudentMember, IEEE*

Kawser Wazed Nafi

Syed Akther Hossain, *Member, IEEE & ACM*

*Abstract*— Cloud Computing is becoming a viable computing solution for services oriented computing. Several open-source cloud solutions are available to these supports. Open-source software stacks offer a huge amount of customizability without huge licensing fees. As a result, open source software are widely used for designing cloud, and private clouds are being built increasingly in the open source way. Numerous contributions have been made by the open-source community related to private-IaaS-cloud. OpenNebula - a cloud platform is one of the popular private cloud management software. However, little has been done to systematically investigate the performance evaluation of this open-source cloud solution in the existing literature. The performance evaluation aids new and existing research, industry and international projects when selecting OpenNebula software to their work. The objective of this paper is to evaluate the load-balancing performance of the OpenNebula cloud management software. For the performance evaluation, the OpenNebula cloud management software is installed and configured as a prototype implementation and tested on the DIU Cloud Lab. In this paper, two set of experiments are conducted to identify the load balancing performance of the OpenNebula cloud management platform- (1) Delete and Add Virtual Machine (VM) from OpenNebula cloud platform; (2) Mapping Physical Hosts to Virtual Machines (VMs) in the OpenNebula cloud platform.

## I. INTRODUCTION

Over the past years, the Cloud phenomenon had an impressive increase in popularity in both the software industry and research worlds. The most interesting feature that Cloud Computing brings, from a Cloud client's point of view, is the on-demand resource provisioning model. This allows Cloud client platforms to be scaled up in order to accommodate more incoming clients and to scale down when the platform has unused resources, and this can all be done while the platform is running. As a result, the physical resources are used more efficiently and the Cloud clients save expenses.

Cloud Computing platforms provide easy accesses to a large pool of computing (and storage) resources through a variety of resources with virtual resource management [2]. Tools and technologies are emerging that can transform an organization existing computing infrastructure into a private or hybrid cloud [3]. Eucalyptus, CloudStack, OpenStack and OpenNebula are the most popular open-source cloud solutions used as to build such community cloud platforms. Eucalyptus is open source cloud software for building AWS-compatible private and hybrid clouds [4]. Apache CloudStack is open source software designed to deploy and manage large networks of virtual machines Infrastructure as a Service (IaaS) cloud computing platform [5]. OpenStack is a cloud operating system that controls large pools of compute, storage, and networking resources throughout a datacenter, all managed through a dashboard that gives administrators control while empowering their users to provision resources through a web interface [6]. OpenNebula is an open-source cloud computing toolkit for managing heterogeneous distributed data center infrastructures. The OpenNebula toolkit manages a data center's virtual infrastructure to build private, public and hybrid implementations of infrastructure as a service [7].

Cloud computing facilities are often supported by virtualization technologies, which enable clouds to acquire or release computing resources on-demand and in a manner such that the loss of any component of the system will no cause system failure [9]. A wide range of visualization solutions have been deployed. KVM [10], Xen [11] and VMware [12] are the three most popular ones. The largest share of cloud adopters plan to use only open-source based, or combination of open-source and VMware based private cloud options. Different visualization techniques may introduce different overhead. Different cloud computing solutions may also introduce different overhead for monitoring and managing the virtual resources [3].

Numerous contributions have been made by the open-source community related to private-IaaS-cloud platforms. These open-source cloud management platforms and tools also have huge impact the adoption of cloud computing technology. OpenNebula is one of the most widely used open-source cloud management software among research institutions and enterprises.

This concept of load balancing is not typical to Cloud platforms and has been around for a long time in the field of distributed systems. In its most abstract form, the problem of load balancing is defined by considering a number of parallel machines and a number of independent tasks, each having its own load and duration [13]. The goal is to assign the tasks to the machines, therefore increasing their load, in such a way as to optimize an objective function. Traditionally, this function is the maximum of the machine loads and the goal is to minimize it. Depending on the source of the tasks, the load balancing problem can be classified as: online load balancing where the set of tasks is known in advance and cannot be modified and online load balancing in the situation that the task set is not known in advance and tasks arrive in the system at arbitrary moments of time.

Load balancing is a process of reassigning the total load to the individual nodes of the collective system to the facilitate networks and resources to improve the response time of the job with maximum throughput in the system [1]. For the performance evaluation of the OpenNebula cloud management software, the following approach is taken. (1) Existing works are reviewed to identify evaluation metrics. (2) Virtual machine management operation is selected for the evaluation. (3) Selective functions, specifically adding and deleting - users and VMs are chosen for the evaluation. (4) Monitoring Host failure and virtual machine failure state load balancing (5) Testing scenarios are identified for the evaluation such as different virtual machine types, number of virtual machines and change in the load of the system, as this reflects an operation environment. (5) The OpenNebula cloud management software, Hypervisors, CentOS (Operative System for the Cloud), OpenNebulaApps (OpenNebulaApps provides a service management layer on top of OpenNebula by configuring the software stack in the applications and managing multi-tiered applications), Haizea (an open-source virtual machine-based lease management architecture) are installed and configured as a prototype implementation and tested on the DIU Cloud Lab.

In this paper, targeting evaluation load balance performance with popular open-source cloud platform - opennebula with four test cases in two groups – (1) add and delete host from opennebula cloud; and (2) mapping physical hosts to virtual machines in 1:4 basic or in the distributed mapping test cases.

## II. EXPERIMENT METHODOLOGY

In this paper, one set of experiments are conducted to benchmark to indentify the load balancing performance of the OpenNebula cloud management platform.

### A. Delete and Add Virtual Machine (VM) from OpenNebula cloud platform:

This experiment evaluates the load balancing performance of opennebula cloud. The result of this experiment can show how opennebula cloud platform cope up with these situation. Within this experiment, we delete a host from virtual network of opennebula cloud and we add a host to virtual network of opennebula cloud – both states are observed to investigate the impact of addition and deduction of VM to and from the virtual network of opennebula cloud in context with online load balance performance.

### B. Mapping Physical Hosts to Virtual Machines (VMs) in the OpenNebula cloud platform:

Within this set of experiments we monitor and evaluate mapping physical hosts to virtual machines in the opennebula cloud. for this experiments, two different test cases: mapping VMs with 1:4 basic physical hosts to virtual machines with virtual network and mapping physical hosts to distributed virtual machines basic in the opennebula sunstone.

## III. EXPERIMENTAL SETUP

### A. Experimental Setup

The OpenNebula cloud management software is installed and configured as a prototype implementation and tested on the DIU Cloud Lab;

Software setup:

- Cloud management platform – OpenNebula
- Hypervisor – KVM
- Operating System – Ubuntu
- OpenNebulaApps
- OpenNebula Sunstone

Hardware setup:

- 64-bit Processor will install and run only on servers with 64-bit x86 CPUs.
- requires a host machine with at least two cores.
- supports only LAHF and SAHF CPU instructions.
- requires the NX/XD bit to be enabled for the CPU in the BIOS.
- supports a broad range of x64 multicore processors.

RAM

- Provide at least 8GB of RAM to take full advantage of opennebula cloud

Hardware Virtualization Support

- To support 64-bit virtual machines, support for hardware virtualization (Intel VT-x or AMD RVI) must be enabled on x64 CPUs.
- To determine whether your server has 64-bit VMware support, download the CPU Identification Utility from vmware.com.

Network Adapters

- One or more Gigabit or 10Gb Ethernet controllers. For a list of supported network adapter models SCSI Adapter, Fibre Channel Adapter or Internal RAID Controller

## IV. EXPERIMENTAL SCENARIOS WITH 5 TEST CASES

Experimental scenarios in with five test cases:

### A. Scenarios for test case 01 and test case 02:

In the OpnenNebula Sunstone, with host management subsystem which has complete functionality for management of physical hosts: create, delete, enable, disable, monitor, list. First we delete one physical host from the opennebula server and after some observation we add one physical host to the opennebula server; and again observe the situation. Create or Delete Host from opennebula sunstone at DIU Cloud Lab showed in figure 1.

We monitor and evaluate in the following two test cases:

*Test case 01:* If we delete a physical host from Subsystem; how opennebula cope up with this situation - with multiple users are connected to the opennebula cloud and in the state they program executions load are high; and, what will happen to the load balance performance for this state.

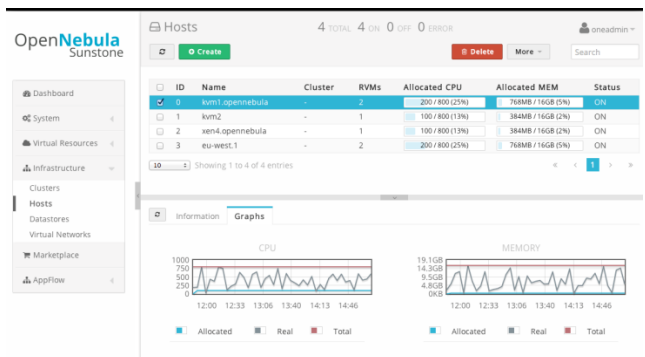

Figure 1 Create or Delete Host from opennebula sunstone at DIU Cloud Lab.

*Test case 02:* If we add a physical host from Subsystem; how opennebula cope up with this situation - with multiple users are connected to the opennebula cloud and in the state they program executions load are high; and, how it will efficiently increase load balance performance for different executions loads in this state.

### B. Scenarios for test case 03 and test case 04:

In order to optimize physical resources by mapping the mapping of virtual resources to physical resources; There is usually a compromise between the following two opposite use cases:

Mapping physical hosts to virtual machines in the opennebula sunstone at DIU Cloud Lab showed in figure 2.

We monitor and evaluate mapping physical hosts to virtual machines in the opennebula sunstone in the following two test cases: (a) test case 03 and (b) test case 04.

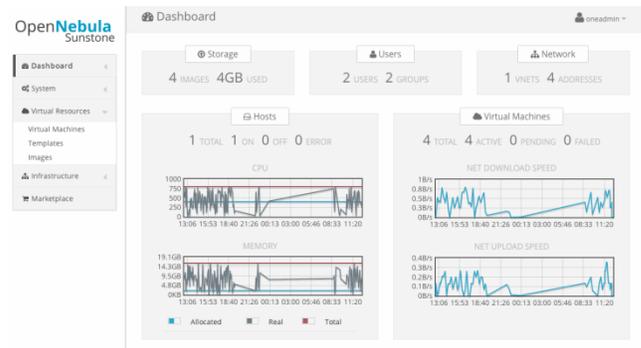

Figure 2 Create or Delete Host from opennebula sunstone at DIU Cloud Lab.

*Test case 03:* Mapping 1:4 basic physical hosts to virtual machines in the opennebula sunstone, showed in figure 5. In the test case, each four virtual machines with singe host connected with two users in active state in two groups; and there are work executions load.

In this scenario, we configure VMs with mapping 1:4 basic physical hosts to virtual machines with virtual network. Each of four VMs is sharing single physical host resources e.g. memories, processors, storages. In this state, our observation

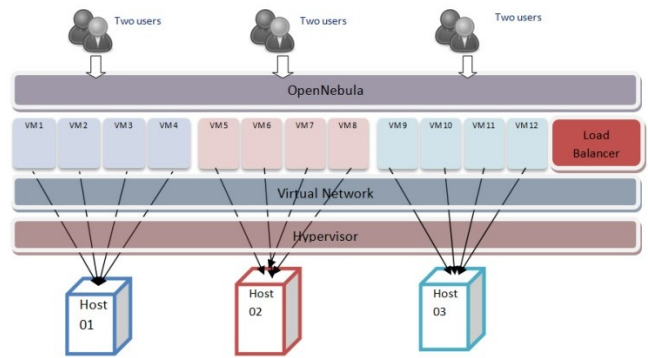

Figure 3 test case 03 state- Mapping 1:4 basic physical hosts to virtual machines at DIU Cloud Lab.

for - if any of physical host fails or power fails or connection fails. To do the physical host fail state, we physically disconnect one host from the network or force power off host machine. In this state, we observe the scenario for one host fails –how the load balancing performance maintain efficiently in this current state of running work execution load from different users.

*Test case 04:* Mapping basic physical hosts to distributed virtual machines in the opennebula sunstone, showed in figure 6. In the test case, each virtual machine with multiple hosts connected with two users in active state in two groups; and there is work execution load.

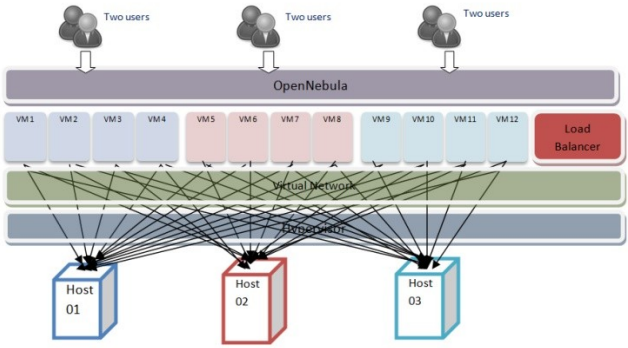

Figure Test case 04 state – Distributed mapping physical hosts to virtual machines at DIU Cloud Lab.

In this scenario, we configure VMs in distributed resource pooling with virtual network. All VMs are sharing physical hosts resources e.g. memories, processors, storages with distributed resource pooling. In this state, our observation for - if any of physical host fails or power fails or connection fails. To do the physical host fail state, we physically disconnect one host from the network or force power off host machine. In this state, we observe the scenario for one host fails –how the load balancing performance maintain efficiently in this current state of running work execution load from different users.

C. *Scenarios for test case 05:*

*Test case 05:* The Cloud provider achieves a more efficient resource usage by trying to minimize the number of physical hosts that are running the virtual resources. In DIU Cloud Lab, numbers of VMs are twelve in a virtual network with 2 physical hosts have been configured. Numbers of users are eight have been logged-in with high loading state. In this state, utilization of maximum physical resources with high numbers of VMs with load balancer for high load from users. This policy is used to minimize the number of physical hosts running in the virtual resources. This state shows in figure 5.

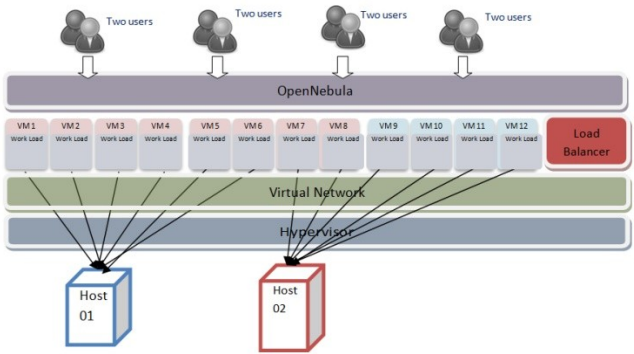

Figure 5: Test case 05 Scenario - minimum the number of physical hosts configured at DIU Cloud Lab.

In this state, our observation for - if any of physical host fails or power fails or connection fails. To do the physical host fail state, we physically disconnect one host from the network or force power off host machine.

V. EXPERIMENTAL RESULTS

A. *Result from test case 01 and test case 02:*

Here numbers of process execution are more or less same within 10-13, from six online users simultaneously request these processes for execution. At the time 40 minutes; we delete one host from the virtual network, and with the same setup on the same time, we add one host to the virtual network. The following different impact happen over the load balancing performance in opennebula cloud platform (observation report showed in figure 6):

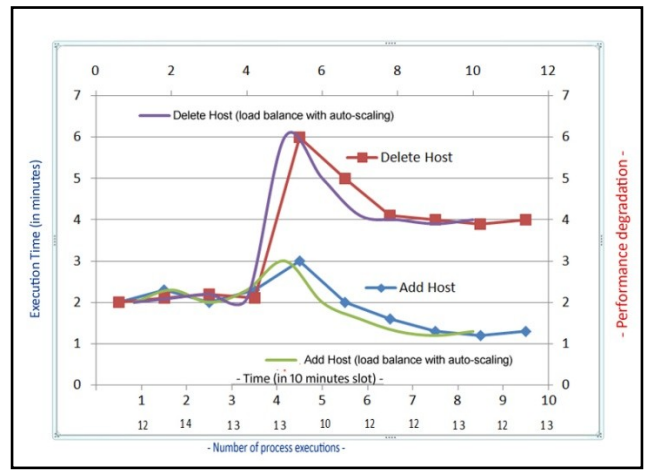

Figure 6: Test case 1 and test case 2 observation result at DIU Cloud Lab.

*Observation from test case 01 and test case 02:*

We observe auto-scaling feature available in the opennebula cloud platform. Load balancing and auto-scaling need to work together in order to get the most efficient platform usage and save expenses. It follows that load balancing should be used in conjunction with auto-scaling efficient performance in order to reduce cost.

*Result from test case 03 and test case 04:*

Here numbers of process execution are more or less same within 12-14, from six online users simultaneously requesting these processes for execution. At the time hours 2; we forced switch off one host from the virtual network, and the following different impact happen over the load balancing performance in opennebula cloud platform (observation report showed in figure 7):

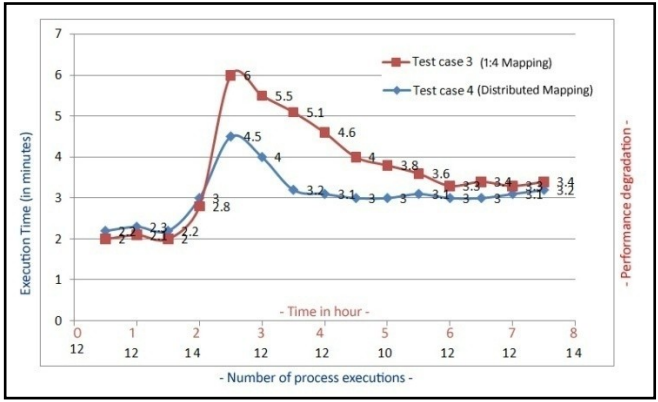

Figure 7: Test case 3 and test case 4 observation result at DIU Cloud Lab.

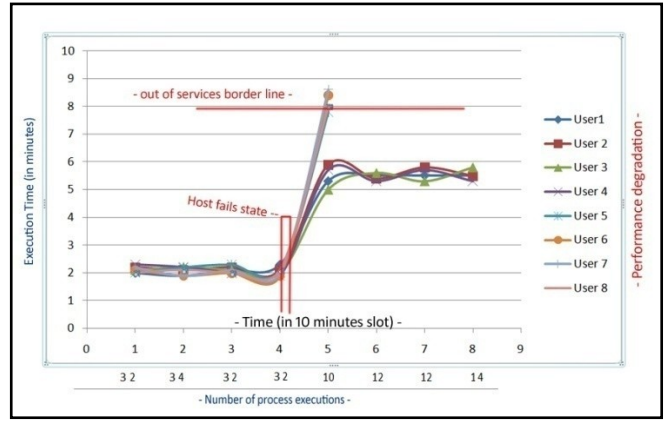

Figure 8: Test case 5 observation result at DIU Cloud Lab.

In the test case 3: Load balance performance degradation is very high compared to test case 4 (distributed mapping) as well as execution time is high accordingly.

In the test case 4: Load balance performance degradation is not very high compared to test case 3 as well as execution time is not high accordingly. Here scale up and scale down the load (execution time and performance) gradually more adjustable faster form compared to test case 3, as all the VMs can execute process.

Observation from the results test case 03 and test case 04: For both test cases investigation results shows - distributed mapping is more efficient for load balancing as well as auto scaling performance in cloud platforms. The virtual resources are distributed across the physical resources. Thus the risk of failure is less for Cloud clients in case of hardware failure. In order to optimize physical resources by mapping the mapping of virtual resources to physical resources; mapping of virtual resources to physical resources also has an impact on Cloud clients.

*Result from test case 05:*

Within the test case 05, utilization of maximum physical resources using high numbers of VMs with load balancer for high load from users. At the time hours 2; we forced switch off one host from the virtual network, and the following different impact happen over the load balancing performance in opennebula cloud platform (observation report showed in figure 8):

In the test case 5: At state the 40 minutes, one physical host down (forced shutdown); Load balance performance degradation is very high as well as execution time is high accordingly. In this state, 50 percentages of users have gone out of services; their process execution requests stack (hold for long time). Other 50 percentage of user's requests stack for a moment and then stat with performance degradation.

Observation from test case 05: For test case investigation results shows - to achieve a more efficient resource usage by trying to minimize the number of physical hosts that are running the virtual resources. The downside for the Cloud client is the fact that his platform is at a greater risk in case of hardware failure because the user's virtual resources are deployed on a small number of physical machines.

## VI. CONCLUSION

In this paper, targeting evaluation load balance performance with popular open-source cloud platform – OpenNebula. For the performance evaluation, existing works are reviewed to identify the metrics, and the OpenNebula cloud management software is installed and configured as a prototype implementation and tested on the DIU Cloud Lab; Experiments are conducted to benchmark to indentify the load balancing performance of the OpenNebula cloud management platform with four test cases in two groups with other one test case – (1) add and delete host from opennebula cloud; and (2) mapping physical hosts to virtual machines in 1:4 basic or in the distributed mapping test cases. To achieve a more efficient resource usage by trying to (3) minimize the number of physical hosts that are running the virtual resources.

In the first group, both test case 03 and test case 04 investigation results shows - auto-scaling feature available in the opennebula cloud platform. Load balancing and auto-scaling need to work together in order to get the most efficient platform usage and save expenses. It follows that load balancing should be used in conjunction with auto-scaling efficient performance in order to reduce cost.

In the second group, both test case 03 and test case 04 investigation results shows - distributed mapping is more efficient for load balancing as well as auto scaling performance in cloud platforms. The virtual resources are distributed across the physical resources. Thus the risk of failure is less for Cloud clients in case of hardware failure. In order to optimize physical resources by mapping the mapping of virtual resources to physical resources; mapping of virtual resources to physical resources also has an impact on Cloud clients.

In the test case 05 investigation results shows - to achieve a more efficient resource usage by trying to minimize the number of physical hosts that are running the virtual resources. The downside for the Cloud client is the fact that his platform is at a greater risk in case of hardware failure because the user's virtual resources are deployed on a small number of physical machines. It follows that minimize the number of physical hosts that are running the virtual resources at a greater risk in case of hardware failure.

## VII. Future work

The evaluation of load balance for any some other popular cloud management platforms e.g. Eucalyptus, OpenStack, CloudStack or Nimbus. The requirements for experimental setups are more or less same for different cloud management platforms; but the result may be varied different CMPs. The test cases may be different for this evaluation for load balancing performance; any suitable test cases may be used for this purpose.

It follows that load balancing should be used in conjunction with auto-scaling in order to reduce cost. It should be evaluated load balancing performance with auto-scaling feature of cloud platform.